\documentclass[nohyper,12pt,letterpaper]{JHEP3}

\usepackage{amsmath}
\usepackage{amsfonts}
\usepackage{lineno}

\usepackage{axodraw4j}
\usepackage{pstricks}
\usepackage{color}
\let\normalcolor\relax

\usepackage{graphicx}

\newcommand{\A}{{\mathcal{A}}}
\newcommand{\bA}{\bar{\mathcal{A}}}
\newcommand{\B}{{\mathcal{B}}}
\newcommand{\bB}{\bar{\mathcal{B}}}

\author{Mat\'ias Leoni and Andrea Mauri \\
Dipartimento di Fisica, Universit\`a di Milano and INFN, Sezione di Milano,
Via Celoria 16, I-20133 Milano, Italy\\
\qquad\\
E-mail: \email{matias.leoni@mi.infn.it, andrea.mauri@mi.infn.it}}

\abstract{We discuss the problem of infrared divergences in the
$\mathcal{N}=2$ superspace approach  to classically marginal
three-dimensional Chern-Simons-matter  theories. Considering the
specific case of ABJM theory, we describe the origin of such
divergences and offer a prescription to eliminate them by
introducing  non-trivial gauge-fixing terms in the action. We also
comment on the extension of our procedure to higher loop order and
to general three-dimensional Chern-Simons-matter models.}

\preprint{IFUM-959-FT}

\title{On the infrared behaviour of 3d Chern-Simons  theories in $\mathcal{N}=2$ superspace}


\begin{document}

\section{Introduction}

The perturbative expansion of the off-shell amplitudes for a
supersymmetric gauge theory may be plagued by infrared divergences
for a number of different motivations. One possible source of
infinities is given by the presence of positive mass-dimension
couplings associated to massless  fields in superrenormalizable
theories.  Canonical Yang-Mills theory coupled to massless matter in
three dimensions turns out to be a good playground to study these
phenomena \cite{Jackiw}. In this case, going high in the order of
the dimensionful coupling in the expansion of a given amplitude,
for dimensional reasons one obtains high powers of  external momenta
in the denominator. If the fields are massless, upon inserting those
amplitudes in higher order graphs, IR divergences inevitably show
up in the Euclidean  integrals.

Considering theories with only dimensionless couplings greatly
improves the situation. A direct power counting argument  shows that
in four-dimensions classical marginality is a sufficient condition
to exclude the presence of IR infinities in amplitudes with generic
external momenta \cite{Poggio}. Even if this argument may be generalized
to marginal three dimensional theories, it only applies to component field formulations and
cannot be applied to superspace formulations as we will show throughout this work.


In fact, a potential source of IR divergences has to be considered as
soon as  the computations are performed using supergraph techniques in supersymmetric gauge theories.
While proving to be an efficient method to compute perturbative
corrections, superspace algebra
comes with additional  infrared issues due to the peculiar nature of
the gauge superfield propagator.  As it is  clearly described in
\cite{GRS}  in the case of four-dimensional Super-Yang-Mills
theories, the appearance of  infrared infinities can be ascribed to
the presence of (corrected) vector lines in loop diagrams. With a
canonically gauge-fixed action and omitting the color structure, the
gauge superfield propagator in momentum space can be written as
\begin{equation}
\frac{1+(\alpha-1)\mathcal{P}_0}{p^2} \,\delta^4(\theta-\theta ')
\end{equation}
where $\alpha$ is the gauge-fixing parameter and
$\mathcal{P}_0=-\frac{1}{p^2}(D^2\bar{D}^2+\bar{D}^2 D^2)$ is the
superspin zero projector. Recall that this operator satisfies
$\mathcal{P}_0+\mathcal{P}_{1/2}=1$, where
$\mathcal{P}_{1/2}=\frac{1}{p^2}D^{\alpha}\bar{D}^2 D_{\alpha}$ is
the superspin $1/2$ projector. It is clear that, already at the
one-loop level,  the Fermi-Feynman gauge $\alpha=1$ is the only
infrared safe choice.  On the other hand radiative corrections,
being governed by Slavnov-Taylor identities, come with the
transverse structure $\mathcal{P}_{1/2}$, thus reintroducing the
infrared dangerous part in the propagator. Therefore, unless a way
is found to perturbatively maintain the Fermi-Feynman form of the
propagator, IR divergences will show up again starting from two-loop
order.

An explicit prescription to cure the IR divergences in the case of
four  dimensional Super-Yang-Mills theory has been given in
\cite{Zanon}. The main idea is  to introduce a non-local  gauge
fixing term and  renormalize  the gauge fixing parameter to preserve
the tree level structure of the Fermi-Feynman propagator. The
prescription presented in \cite{Zanon} is strongly based on the
nature of the model (gauge sector of Yang-Mills type) and on the
space-time dimension.

The aim of this paper is to study the infrared behaviour of
supergraph amplitudes in the  case of marginal Chern-Simons-matter
systems in three dimensions described in the $\mathcal{N}=2$
superspace formalism.  These models can be treated  in strong
analogy with 4d Yang-Mills theories  while exhibiting a completely
different gauge structure. Due to the recent interest in CS-matter
models in the context of AdS/CFT  we start our analysis in the
specific case of ABJM theory \cite{ABJM}. By directly computing
Green's functions up to two-loop order,  we show that infrared
divergences appear in the amplitudes but can be seen as a gauge
artefact of the formalism. At first, we gauge fix the ABJM action in
a canonical way ($\alpha$-gauge) and show that, in analogy with the
four-dimensional SYM case, IR infinities show up when a corrected
gauge vector propagator is inserted in loop amplitudes. By direct
inspection of the dependence of the infrared singularities on the
gauge fixing parameter $\alpha$,  we conclude that there is no
suitable choice for the latter that both eliminate the divergences
and preserve hermiticity of the action.

To solve this problem, we slightly revise the prescription of
\cite{Zanon} introducing a set of non-canonical gauge fixing terms
($\eta$-gauge). This new set of gauges has the virtue that it can be
used, by perturbatively fine tuning the parameter $\eta$, to
complete loop by loop the transverse structure of the gauge vector
propagator with the longitudinal part, thus improving its behaviour
in the infrared. We will show how the infinities are canceled in
this way by direct perturbative computations. The $\eta$-gauge can
hence be considered as a tool to consistently study the perturbative
expansion of the amplitudes of the model without the presence of IR
divergences. It's important to stress that infrared
divergences, being an artefact of the superspace formalism, do not
manifest themself in physical gauge invariant quantities. In this case
the $\alpha$- and $\eta$-gauges produce coincident results.

As a byproduct of our analysis, we explicitly compute the finite
expression in a general gauge for the two-loop propagator of the
chiral superfield in ABJM. Moreover, we study a special vanishing
external momenta limit of the two-loop vertex function of ABJM
theory showing that it produces a finite result. Finally we comment
on the extension of our results to a general perturbative order and
to any classically marginal Chern-Simons-matter system.

\section{ABJM action and gauge-fixing}

To address the problem of IR divergences in three-dimensional
Chern-Simons theories we restrict ourself  to the specific case of
ABJM model. This theory possesses remarkable properties such as
extended supersymmetry and exact conformal invariance which will
simplify the analysis of the infrared behaviour. We  extend our
results to more general CS theories in Section
\ref{section_all_loop}.

At first, we set up our notations and quantize the theory
introducing the standard gauge fixing procedure and an alternative
one that will ensure the cancelation of the IR divergences in loop
amplitudes. We will use the $\mathcal{N}=2$ superspace formulation
first presented in \cite{Klebanov} adapted to the notations of
\cite{Superspace} (see Appendix A for further details).  In
Euclidean space, we quantize the theory with a path integral measure
of the form $\int D\phi\ e^{S[\phi]}$.  ABJM theory can then be
written as  $\mathcal S=\mathcal S_{CS}+\mathcal S_{mat}+\mathcal
S_{pot}$, where
\begin{align}\label{action1}
& \mathcal S_{CS}=\frac{k}{4\pi}\int d^3x d^4\theta\int\limits_{0}^{1}dt\
\mathrm{tr} \left[V\bar{D}^\alpha\left(e^{-tV}D_{\alpha}
e^{tV}\right)-\hat V\bar{D}^\alpha\left(e^{-t\hat V}D_{\alpha}
e^{t\hat V}\right)\right] \\
& \mathcal S_{mat}=\int d^3x d^4\theta\ \mathrm{tr}\left(\bA^A e^{\hat V}\A_A
e^{-V}+\bB_A e^{V}\B^A e^{-\hat V}\right) \label{action2} \\
&\mathcal S_{pot}=\frac{2\pi i}{k}\int d^3x d^2\theta\, \epsilon_{AC}\epsilon^{BD}
\mathrm{tr}(\B^A\A_D\B^C\A_B)+\frac{2\pi i}{k}\int d^3x d^2\bar\theta\,
\epsilon^{AC}\epsilon_{BD} \mathrm{tr}(\bB_A\bA^D\bB_C\bA^B). \label{action3}
\end{align}
The chiral superfields $\B^A$ and $\A_A$ (where
$\scriptstyle{A,B,C,D=1,2}$) transform in the $(\mathbf 2,\mathbf
1)$ and $(\mathbf 1,\mathbf 2)$ of the global $SU(2)\times SU(2)$
respectively. Moreover, they transform in the $(\mathbf
N,\bar{\mathbf N})$ and $(\bar{\mathbf N},\mathbf N)$ of the gauge
group $U_k(N)\times U_{-k}(N)$, such that if explicit gauge group
labeling is needed, the chiral superfields are $\B^a_{\hat{a}}$,
$\bB_a^{\hat{a}}$, $\A_a^{\hat{a}}$, $\bA^a_{\hat{a}}$, with
$a,b,\hat{a},\hat{b}=1,..,N$. The gauge vector superfields $V$ and
$\bar V$ are in the adjoint representation of the groups and may be
written either as $V=T^i\,V_i$ with $i=1,\cdots,N^2$ or with matrix
labeling $V^a_b$, $\hat V_{\hat a}^{\hat b}$.

To quantize the theory we re-scale the vector gauge fields
$V\rightarrow\sqrt{\frac{4\pi}{k}}V$ and we choose in each gauge
sector the gauge fixing functions $F=\bar D^2 V$, $\bar F=D^2 V$.
The standard procedure in $d=3$ is to introduce in the functional
integral the factor:
\begin{equation}
\int\mathcal{D}f\,\mathcal{D}\bar f\, \Delta(V)\Delta^{-1}(V)
\exp\left(\frac{1}{2\alpha}\int d^3x\,d^2\theta\,\mathrm{tr}\left(f\,f\right)\right)
\exp\left(\frac{1}{2\bar\alpha}\int d^3x\,d^2\bar\theta\,\mathrm{tr}\left(\bar f\,\bar f\right)\right),
\end{equation}
where
\begin{equation}
\Delta(V)=\int d\Lambda\, d\bar\Lambda\,\delta(F(V,\Lambda\,\bar\Lambda)-f)\,\delta(\bar{F}(V,\Lambda\,\bar\Lambda)-\bar{f}),
\end{equation}
with $\Lambda$ the chiral superfield of the gauge transformation
$e^V\rightarrow e^{i\bar\Lambda}\,e^V\,e^{-i\Lambda}$ and $\alpha$ a
dimensionless parameter. Notice that, following \cite{Penati}, we
are introducing a gauge averaging given by gaussian weights with
chiral integrals of the form $\sim e^{\int ff} e^{\int \bar f\bar
f}$. This is to be contrasted with the standard $d=4$,
$\mathcal{N}=1$ superspace procedure where one averages with a
non-chiral (whole superspace) gaussian weight of the form $e^{\int
\bar f f}$. The average produces the canonical quadratic gauge fixed
action \cite{Kazakov}
\begin{align}
   \mathcal S_{gf}^{(\alpha)}=\frac{1}{2}\int d^3x d^4\theta\ \mathrm{tr}&
   \left[ V\left(\bar{D}^{\gamma}D_{\gamma}+\frac{1}{\bar\alpha} D^2+\frac{1}{\alpha}\bar{D}^2\right)V\right] \nonumber \\
& - \mathrm{tr}\left[ \hat V\left(\bar{D}^{\gamma}D_{\gamma}+\frac{1}{\bar\alpha} D^2+\frac{1}{\alpha}\bar{D}^2\right)\hat V
\right],
\end{align}
such that, after inverting the kinetic operator we obtain the gauge field propagators in momentum space
\begin{equation}\label{trivialpropagators}
  \begin{picture}(0,0) (125,-72)
    \SetWidth{1.0}
    \SetColor{Black}
    \Photon(80,-71)(112,-71){3.5}{3}
    \Text(64,-71)[lb]{\small{\Black{$V_{\ b}^a$}}}
    \Text(112,-71)[lb]{\small{\Black{$V_{\ d}^c$}}}
  \end{picture}
=\frac{1}{p^2}\left(\bar D^{\alpha} D_{\alpha}+\alpha D^2 +\bar\alpha\bar D^2\right)\delta^4_{(\theta,\theta')}\delta^b_c\delta^d_a,
\nonumber
\end{equation}
\begin{equation}
\begin{picture}(0,0) (125,-70)
    \SetWidth{1.0}
    \SetColor{Black}
    \Photon(80,-71)(112,-71){3.5}{3}
    \Text(64,-71)[lb]{\small{\Black{$\hat V_{\ \hat b}^{\hat a}$}}}
    \Text(112,-71)[lb]{\small{\Black{$\hat V_{\ \hat d}^{\hat c}$}}}
  \end{picture}
=-\frac{1}{p^2}\left(\bar D^{\alpha} D_{\alpha}+\alpha D^2 +\bar\alpha\bar D^2\right)\delta^4_{(\theta,\theta')}\delta^{\hat b}_{\hat c}\delta^{\hat d}_{\hat a},
\end{equation}
with $\delta^4_{(\theta,\theta')}=\delta^4(\theta-\theta')$. From
now on we shall call this gauge fixing procedure as the
``$\alpha$-gauge''. This propagator simplifies greatly when
$\alpha\rightarrow 0$ (Landau Gauge). In the next Section we will
see that, if we want to preserve the hermiticity of the gauge fixed
action considering $\alpha$ and $\bar{\alpha}$ as complex
conjugates, then the infrared divergences cannot be canceled by a
simple fine tuning of  the gauge fixing parameter.

To solve this problem, in analogy with \cite{Zanon}, we propose a
different gauge averaging procedure. We choose the same gauge fixing
functions as before, but this time we introduce the following term
in the functional integral:
\begin{equation}\label{gaussian}
\mathrm{det}\mathcal{\hat M}
\int \mathcal{D}f\,\mathcal{D}\bar f\, \Delta(V)\Delta^{-1}(V)
\exp\left(\int d^3x\,d^4\theta\,\mathrm{tr}\left(\bar f\hat{\mathcal{M}} f\right)\right).
\end{equation}
In this way we allow for a non-trivial gauge averaging by the
insertion of the operator $\hat{\mathcal{M}}$. It is important to
stress that as long as $\hat{\mathcal{M}}$ is field independent, the
$\mathrm{det}\hat{\mathcal{M}}$ factor appearing in the functional
integral is irrelevant and there is no need to introduce
Nielsen-Kallosh ghosts \cite{NK} in the action. In $d=4$ the
$\hat{\mathcal{M}}$ operator is dimensionless and one can simply
choose $\hat{\mathcal{M}}=\hbox{constant}$. In three dimensions it
has dimensions of length so that we may choose either a dimensionful
constant or a non-local gauge fixing term. Our choice of this
operator in momentum space is
$\hat{\mathcal{M}}(p)=\tfrac{1}{\eta^{\epsilon}_{(p)}\,|p|}$, where
$\eta^{\epsilon}_{(p)}$ is a dimensionless function that contains
$\epsilon$ powers of $p$ \footnote{This choice would introduce
Nielsen-Kallosh ghosts in the action if the computations were
performed using the background field method as in \cite{Zanon}.}.
The early introduction of the $\epsilon=\tfrac{3}{2}-\tfrac{d}{2}$
regulator parameter has to be understood formally in the sense of
dimensional reduction, that is, we will still perform D-algebra
calculations in three dimensions and only at the end we will
regularize Feynman integrals. More specifically, we will define
$\eta$ as an odd power series in the 't Hoof coupling
$\lambda=\tfrac{N}{k}$ with coefficients that we will conveniently
choose. By choosing the same Gaussian measure (\ref{gaussian}) for
both gauge sectors  we obtain the gauge fixed action in momentum
space
\begin{align}
   \mathcal S_{gf}^{(\eta)}=\frac{1}{2}\int\frac{d^3p}{(2\pi)^3} d^4\theta\ \mathrm{tr}&
   \left[ V(-p)\left(\bar{D}^{\gamma}D_{\gamma}-\frac{|p|}{\eta^{\epsilon}_{(p)}} \mathcal{P}_0\right)V(p)\right] \nonumber \\
& - \mathrm{tr}\left[ \hat V(-p)\left(\bar{D}^{\gamma}D_{\gamma}+\frac{|p|}{\eta^{\epsilon}_{(p)}} \mathcal{P}_0\right)\hat V(p)
\right].
\end{align}
Inverting the operators of the quadratic part of the gauge fixed action we obtain the gauge field propagators
\begin{equation}
  \begin{picture}(0,0) (125,-72)
    \SetWidth{1.0}
    \SetColor{Black}
    \Photon(80,-71)(112,-71){3.5}{3}
    \Text(64,-71)[lb]{\small{\Black{$V_{\ b}^a$}}}
    \Text(112,-71)[lb]{\small{\Black{$V_{\ d}^c$}}}
  \end{picture}
=\left(\frac{\bar D^{\alpha} D_{\alpha}}{p^2}+\frac{\eta^{\epsilon}_{(p)}}{|p|}\mathcal{P}_0\right) \delta^4_{(\theta,\theta')}\delta^b_c\delta^d_a,
\nonumber
\end{equation}
\begin{equation}
\begin{picture}(0,0) (125,-70)
    \SetWidth{1.0}
    \SetColor{Black}
    \Photon(80,-71)(112,-71){3.5}{3}
    \Text(64,-71)[lb]{\small{\Black{$\hat V_{\ \hat b}^{\hat a}$}}}
    \Text(112,-71)[lb]{\small{\Black{$\hat V_{\ \hat d}^{\hat c}$}}}
  \end{picture}
=\left(-\frac{\bar D^{\alpha} D_{\alpha}}{p^2}+\frac{\eta^{\epsilon}_{(p)}}{|p|}\mathcal{P}_0\right) \delta^4_{(\theta,\theta')}\delta^b_c\delta^d_a.
\end{equation}
We shall call this gauge fixing procedure as the ``$\eta$-gauge''.
We will show that  allowing  $\eta$ to be corrected order by order
in the 't Hooft coupling  $\frac{N}{k}$, we will be able to cancel
the infrared divergent parts in the amplitudes. To complete the
gauge fixing procedure, we rewrite the $\Delta^{-1}(V)$ factor  in
the path integral  using  the usual Fadeev-Popov $b$-$c$ and $\hat
b$-$\hat c$ system of Grassmanian chiral superfield ghosts. In both
of our gauge fixing choices the ghost action reads
\begin{align}\label{fp_action}
\mathcal S_{fp}=\int d^3x d^4\theta\ \mathrm{tr}&
\Big[\bar b\,c+\bar c\,b+\frac{1}{2}\sqrt{\frac{4\pi}{k}}(b+\bar b)[V,c+\bar c]\nonumber \\
&- \hat{\bar b}\,\hat{c}- \hat{\bar c}\,\hat{b}-
  \frac{1}{2}\sqrt{\frac{4\pi}{k}}(\hat{b}+\hat{\bar b})[\hat{V},\hat{c}+\hat{\bar c}]
 +\mathcal O (1/k)\Big].
\end{align}
In Appendix B we detail some of the relevant Feynman rules for ABJM theory.

\section{Infrared behaviour of amplitudes in ABJM theory}

We would like now to understand the origin of the IR divergences in
the perturbative expansion of the off-shell amplitudes.  In order to
do so, we  compute the one-loop correction to the vector gauge
superfield $V$ in the $\alpha$- and $\eta$-gauges. Performing a
direct two-loop computation of the (finite) corrections to the
self-energy of the matter superfield and to the superpotential, we
will see that  IR infinities only arise when the one-loop corrected
gauge propagator is inserted in loop diagrams. A suitable choice of
the $\eta$ gauge fixing parameter will then cancel  the divergences.
Our explicit examples will be completed with an all loop analysis in
Section \ref{section_all_loop}.

\subsection{One-loop vector propagator}
\begin{figure}[h!]
    \centering
    \includegraphics[width=0.5\textwidth]{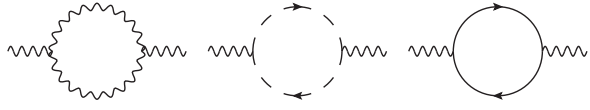}
    \caption{One-loop self-energy corrections of the gauge vector superfields.}
    \label{self energy gauge}
\end{figure}

The one loop corrected gauge vector field receives contributions
from matter, ghost and gauge vector fields as we show in Figure
\ref{self energy gauge}.
In the $\alpha$-gauge these evaluate
\begin{align}
&\Delta^{(1)}_{V-\scriptsize{gauge}}=\frac{\pi}{k}(\delta^b_c\delta^d_a N-\delta^b_a\delta^d_c)\int\frac{d^3k\,d^4\theta}{(2\pi)^3}G^{\epsilon}_{(1,1)}(k^2)^{1/2-\epsilon}\,
V^a_{\ b}(-k)\left(\mathcal{P}_{0}+\alpha\bar\alpha\,\mathcal{P}_{1/2}\right)V^c_{\ d}(k)\nonumber\\
&\Delta^{(1)}_{V-\scriptsize{ghost}}=-\frac{\pi}{k}(\delta^b_c\delta^d_a N-\delta^b_a\delta^d_c)\int\frac{d^3k\,d^4\theta}{(2\pi)^3}G^{\epsilon}_{(1,1)}(k^2)^{1/2-\epsilon}\,
V^a_{\ b}(-k)\left(\mathcal{P}_{0}+\mathcal{P}_{1/2}\right)V^c_{\ d}(k)\nonumber\\
&\Delta^{(1)}_{V-\scriptsize{matter}}=\frac{4\pi}{k}\delta^b_c\delta^d_a N\int\frac{d^3k\,d^4\theta}{(2\pi)^3}G^{\epsilon}_{(1,1)}(k^2)^{1/2-\epsilon}\,
V^a_{\ b}(-k)\mathcal{P}_{1/2}V^c_{\ d}(k)\nonumber\\
&\Delta^{(1)}_{\scriptsize{mixed}}=-\frac{4\pi}{k}\delta^b_c\delta^{\hat
d}_{\hat a}
\int\frac{d^3k\,d^4\theta}{(2\pi)^3}G^{\epsilon}_{(1,1)}(k^2)^{1/2-\epsilon}\,
\hat{V}^{\hat a}_{\ \hat d}(-k)\mathcal{P}_{1/2}V^c_{\ b}(k)\ .
\end{align}
Here we are displaying the corrections to the $V-V$ propagator as
well as to the mixed $V-\hat V$ propagator (last line). The latter
receives contributions only from the matter diagram of Figure
\ref{self energy gauge}. The corrections to the $\hat V-\hat V$
propagator can be easily read from the $V-V$ case. The definition of
the $G^{\epsilon}_{(a,b)}$ functions can be found in Appendix C.
Notice that the mixed $V-\hat V$ contribution is subleading in N.
Consistently with the Slavnov-Taylor identities, the complete
one-loop correction to $V-V$ contains only the spin $1/2$ projection
of the gauge field and is given by
\begin{equation}\label{complete_gauge_vector}
\Delta^{(1)}_{V}=\frac{\pi}{k}\left((3+\bar\alpha\alpha)\delta^b_c\delta^d_a N+(1-\bar\alpha\alpha)\delta^b_a\delta^d_c\right)\int\frac{d^3k\,d^4\theta}{(2\pi)^3}G^{\epsilon}_{(1,1)}(k^2)^{1/2-\epsilon}\,
V^a_{\ b}(-k)\mathcal{P}_{1/2}V^c_{\ d}(k).
\end{equation}
With the particular choice $\alpha=0$ (Landau gauge) we exactly
reproduce the results of \cite{Penati}\cite{ASM}. We notice that,
looking at the leading part of the correction, the effect of working
in a general $\alpha$-gauge simply results in a  positive constant
shift $|\alpha|^2$. Therefore we conclude that, if we want to
preserve hermiticity of the action, there is no way to fully
eliminate the one-loop correction by fine tuning the gauge-fixing
parameter.

Now we work out the correction in the $\eta$-gauge restricting the
analysis to leading\footnote{The following ideas may be also worked
out at subleading order in $N$ but we would have to add a mixed
$\bar V-V$ gauge fixing term.} order in $N$. As anticipated, the
idea is to make a perturbative expansion of the gauge-fixing
parameter in the 't Hooft coupling
\begin{equation}\label{expo}
\eta^{\epsilon}_{(k)}=\,^1\eta^{\epsilon}_{(k)}\lambda+O(\lambda^3),
\end{equation}
such that to order $\lambda$ we obtain the total correction of the propagator:
\begin{equation}
\lambda\,\frac{(-6\pi G^{\epsilon}_{(1,1)}(k^2)^{-\epsilon}\mathcal{P}_{1/2}+ \,^1\eta^{\epsilon}_{(k)}\mathcal{P}_{0})}{(k^2)^{1/2}}\delta_{(\theta,\theta')}.
\end{equation}
It's easy to see that  if we choose $\,^1\eta^{\epsilon}_{(k)}=-6\pi G^{\epsilon}_{(1,1)}(k^2)^{-\epsilon}$  we exactly complete the transverse structure $\mathcal{P}_{1/2}$ with the longitudinal part $\mathcal{P}_0$ to obtain
\begin{equation}\label{eta_corrected_propagator}
-6\pi\,G^{\epsilon}_{(1,1)}\lambda\frac{\delta_{(\theta,\theta')}}{(k^2)^{1/2+\epsilon}}.
\end{equation}
In the next Section we compute two-loop Green's functions and show
that the improved IR behaviour of the $\eta$-gauge propagator in
(\ref{eta_corrected_propagator}) is enough to cure the problem of IR
infinities.

\subsection{Matter  self-energy at two-loop order}

\subsubsection{Landau gauge}

Working in the Landau gauge simplifies greatly the calculation since
many diagrams can be discarded due to the form of the gauge vector
propagator. It is easy to see that all one-loop corrections vanish
with standard gauge averaging. In Figure \ref{self energy
corrections} we display all non-vanishing self energy two loop
quantum corrections of matter fields in this gauge. The blob in
diagrams (c) and (d) represents the insertion of the full one loop
correction to the gauge propagator. Any other potentially
contributing diagram is zero either by D-algebra or by color
symmetry.

\begin{figure}[h!]
    \centering
    \includegraphics[width=0.5\textwidth]{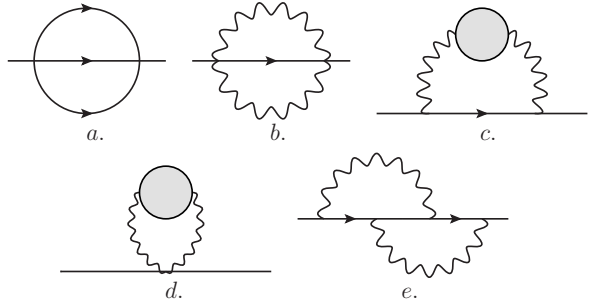}
    \caption{Two loop self-energy quantum corrections.}
    \label{self energy corrections}
\end{figure}

Let us for example calculate with detail diagram $b$ of Figure
\ref{self energy corrections}. Taking into account the possibility
of having $V$-$\hat V$, $V$-$V$ and $\hat V$-$\hat V$ internal lines
and using color vertex factors as (\ref{2vectormatter}), it
evaluates to
\begin{equation}
\Pi_b=-\frac{1}{2}\left(N^2-1\right)\left(\frac{4\pi}{k}\right)^2\int d^4\theta\ d^4\theta'\int \frac{d^3p}{(2\pi)^3}\ \mathrm{tr}\left(\bB_A(-p)\B^A(p)\right)\ \mathcal{D}_b(\theta,\theta')
\end{equation}
with
\begin{equation}
\mathcal{D}_b(\theta,\theta')=\int \frac{d^3 k\,d^3 l}{(2\pi)^3 (2\pi)^3}\ \frac{\bar D^{\alpha}D_{\alpha}\delta^4_{(\theta,\theta')} \bar D^2 D^2\delta^4_{(\theta,\theta')} \bar D^{\beta}D_{\beta}\delta^4_{(\theta,\theta')}}{l^2\,k^2\,(k+l+p)^2},
\end{equation}
the D-algebra factor of the supergraph. As we mentioned before, we
perform all D-algebra manipulations in three dimensions and we
calculate the final Feynman integral in d dimensions. After the
usual integration by parts we obtain an ultraviolet divergent
contribution
\begin{equation}
\Pi_b=\left(N^2-1\right)\left(\frac{4\pi}{k}\right)^2\int \frac{d^3p\,d^4\theta}{(2\pi)^3}\ \mathrm{tr}\left(\bB_A(-p)\B^A(p)\right)\ G^{\epsilon}_{(1,1)}G^{\epsilon}_{\left(1,1/2+\epsilon\right)}\,(p^2)^{-2\epsilon}.
\end{equation}

To obtain the contribution $a$ from Figure \ref{self energy
corrections} we need vertex factors (\ref{colorsuperpotential1}) and
(\ref{colorsuperpotential2}). We get the UV divergent contribution
$\Pi_a=2\Pi_b$.

Using the corrected vector propagator we obtain for graph $d$ of
Figure \ref{self energy corrections} an UV/IR divergent tadpole
\begin{equation}
\Pi_d=-3\left(N^2-1\right)\left(\frac{4\pi}{k}\right)^2\int \frac{d^3p\,d^4\theta}{(2\pi)^3}\ \mathrm{tr}\left(\bB_A(-p)\B^A(p)\right)\ G^{\epsilon}_{(1,1)}\int\frac{d^d k}{(2\pi)^d}\frac{1}{(k^2)^{\frac{3}{2}+\epsilon}}.
\end{equation}
and for graph $c$ an infrared divergent contribution
\begin{equation}
\Pi_c=3\left(N^2-1\right)\left(\frac{4\pi}{k}\right)^2\int \frac{d^3p\,d^4\theta}{(2\pi)^3}\ \mathrm{tr}\left(\bB_A(-p)\B^A(p)\right)\ G^{\epsilon}_{(1,1)}\int\frac{d^d k}{(2\pi)^d}\frac{2p.(p+k)}{(k^2)^{\frac{3}{2}+\epsilon}(k+p)^2}.
\end{equation}
Summing up $a$, $b$, $c$ and $d$ we obtain the cancelation of all UV
divergent contributions and we are left with an IR divergent piece
\begin{equation}
\Pi_{a+b+c+d}=3\left(N^2-1\right)\left(\frac{4\pi}{k}\right)^2\int \frac{d^3p\,d^4\theta}{(2\pi)^3}\ \mathrm{tr}\left(\bB_A(-p)\B^A(p)\right)\ \mathcal G_d(p),
\end{equation}
with $\mathcal G_d(p)$ as given in the appendix. Finally, diagram
$e$ produces a finite correction
\begin{equation}
\Pi_{e}=-2\left(N^2-1\right)\left(\frac{4\pi}{k}\right)^2\int \frac{d^3p\,d^4\theta}{(2\pi)^3}\ \mathrm{tr}\left(\bB_A(-p)\B^A(p)\right)\ I_e
\end{equation}
where $I_e$ is the factor obtained after closing the D-Algebra:
\begin{equation}
I_e=\int \frac{d^3 k\,d^3 l}{(2\pi)^3 (2\pi)^3}\ \frac{(k+p)^2 (l+p)^2-k^2 l^2+p^2 (k+l+p)^2}{k^2\,(k+p)^2\,(k+l+p)^2\,l^2\,(l+p)^2}=\frac{1}{64}.
\end{equation}
To conclude, the sum of all contributions gives a finite and an
infrared divergent piece
\begin{equation}
\Pi=\left(N^2-1\right)\left(\frac{4\pi}{k}\right)^2\int \frac{d^3p\,d^4\theta}{(2\pi)^3}\ \mathrm{tr}\left(\bB_A(-p)\B^A(p)\right)\ \left(3\mathcal G_d(p)-\frac{1}{32}\right).
\end{equation}
Working in the Landau gauge, we explicitly see that  infrared
infinities are only given by graphs c and d, which correspond to
insertion of the 1-loop corrected vector propagator.

\subsubsection{$\alpha$-gauge}

\begin{figure}[h!]
    \centering
    \scalebox{1.5}{
    \includegraphics[width=0.5\textwidth]{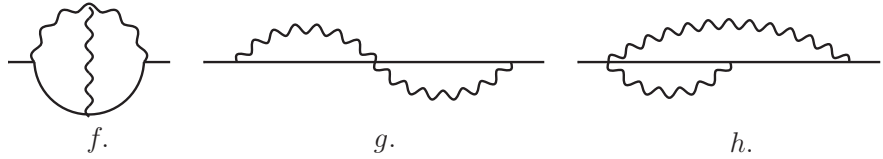}}
    \caption{Additional quantum corrections in the $\alpha$-gauge.}\label{additional}
\end{figure}

We now take the more general case $\alpha\ne 0$. Once again  there
are no one-loop matter corrections. The list of two-loop self energy
contributions gets larger. Apart from those already displayed in
Figure \ref{self energy corrections}, which are modified by the more
general $\alpha$-dependent propagator, we have some additional
contributions displayed in Figure \ref{additional}.
The new contributions produce additional UV, IR divergences and
finite pieces. The sum of the original diagrams we had, with the
modified $\alpha$-dependent propagator gives
\begin{align}
\Pi^{\alpha}_{a+\cdots+e}=&\left(N^2-1\right)\left(\frac{4\pi}{k}\right)^2 \int \frac{d^3p\,d^4\theta}{(2\pi)^3}\ \mathrm{tr}\left(\bB_A(-p)\B^A(p)\right)\nonumber\\ &\times\left(-4\alpha\bar\alpha\, G^{\epsilon}_{(1,1)}G^{\epsilon}_{(1,1/2+\epsilon)}\,(p^2)^{-2\epsilon}+(3+\alpha\bar\alpha)\mathcal G_d(p)-\frac{1}{32}(1+\alpha\bar\alpha)\right).
\end{align}
And the contributions from the additional diagrams of Figure \ref{additional}
\begin{align}
\Pi^{\alpha}_{f+g+h}=&\left(N^2-1\right)\left(\frac{4\pi}{k}\right)^2 \int \frac{d^3p\,d^4\theta}{(2\pi)^3}\ \mathrm{tr}\left(\bB_A(-p)\B^A(p)\right)\nonumber\\ &\times\left(4\alpha\bar\alpha\,G^{\epsilon}_{(1,1)}G^{\epsilon}_{(1,1/2+\epsilon)}\, (p^2)^{-2\epsilon}+\frac{1}{32}\alpha\bar\alpha\right).
\end{align}
By summing up all the contributions, we find as expected that all UV
$\alpha$-dependent divergences cancel out. The finite piece we had
already encountered in the Landau gauge is not modified, and the IR
divergent piece gets shifted:
\begin{equation}
\Pi^{\alpha}=\left(N^2-1\right)\left(\frac{4\pi}{k}\right)^2\int \frac{d^3p\,d^4\theta}{(2\pi)^3}\ \mathrm{tr}\left(\bB_A(-p)\B^A(p)\right)\ \left((3+\alpha\bar\alpha)\mathcal G_d(p)-\frac{1}{32}\right).
\end{equation}
From this we may conclude that if we only allow a hermitian
gauge-fixed action, such that $\bar\alpha$ is literally the complex
conjugate of $\alpha$, then it is not possible to choose a value of
the gauge fixing parameter $\alpha$ such that the self-energy
corrections are infra-red safe. This is a direct consequence of the
fact that IR divergences are eventually produced only by corrected
vector propagators. It's also important to notice that  the finite
correction to the propagator turns out to be gauge independent, even
if the propagator itself is not a physical quantity.

\subsubsection{$\eta$-gauge}

\begin{figure}[h!]
    \centering
    \includegraphics[width=0.5\textwidth]{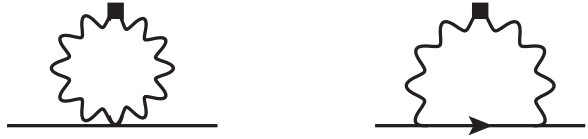}
    \caption{One loop corrections in the $\eta$-gauge. The small black squares in the gauge vector propagators should be intended as the $\eta$ dependent piece of the propagator.}\label{1loop}
\end{figure}

In the $\eta$-gauge the vector superfield propagator is written as:
\begin{equation}
\left<V^c_d(-p)\,V^a_b(p)\right>=\left(\frac{\bar D^{\alpha} D_{\alpha}}{p^2}+\frac{\eta^{\epsilon}_{(p)}}{|p|} \mathcal{P}_0\right)\delta^4_{(\theta,\theta')}\delta^b_c\delta^d_a.
\end{equation}
where $\eta^{\epsilon}_{(p)}$ is  expanded as  in (\ref{expo}). The
first piece of the propagator gives rise to matter self energy
diagrams starting from two loops with the same contributions as in
the Landau gauge (see fig. \ref{self energy corrections}) such that,
for $N\gg1$, it gives a finite and an infrared divergent piece of
order $(\frac{N}{k})^2$ given by
\begin{equation}
\Pi^{\eta}_{a+\cdots+e}=\left(\frac{4\pi N}{k}\right)^2\int \frac{d^3p\,d^4\theta}{(2\pi)^3}\ \mathrm{tr}\left(\bB_A(-p)\B^A(p)\right)\ \left(3\mathcal G_d(p)-\frac{1}{32}\right).
\end{equation}
The second part of the propagator produces one-loop corrections to
matter self energy such that, if the gauge parameter is of order
$\frac{N}{k}$, the contribution is of order $(\frac{N}{k})^2$. In
this case, two loop and higher corrections will contribute beyond
$(\frac{N}{k})^2$ so we do not consider them. The one-loop
contributions are displayed in Figure \ref{1loop}.

\noindent After straightforward D-algebra, and with the choice
$\eta^{\epsilon}_{(k)}=-6\pi
G^{\epsilon}_{(1,1)}(k^2)^{-\epsilon}\lambda+O(\lambda^3)$ we obtain
\begin{equation}
\Pi^{\eta}_{\scriptsize\mbox{1-loop}}=-\left(\frac{4\pi N}{k}\right)^2\int \frac{d^3p\,d^4\theta}{(2\pi)^3}\ \mathrm{tr}\left(\bB_A(-p)\B^A(p)\right)3\mathcal{G}_d(p).
\end{equation}
From this we see that with the choice of $\eta^{\epsilon}_{(p)}$
that produces the IR improved gauge propagator, we cancel the
infrared divergent part obtaining only the universal finite piece
already computed in the $\alpha$-gauge:
\begin{equation}\label{finalSelfEnergy}
\Pi^{\eta}=\Pi^{\eta}_{a+\cdots+e}+\Pi^{\eta}_{\scriptsize\mbox{1-loop}}=-\frac{1}{2}\pi^2 \frac{N^2}{k^2}\int d^3x\,d^4\theta\ \mathrm{tr}\left(\bB_A\B^A\right).
\end{equation}
We therefore conclude that the improved  IR behaviour of the gauge
propagator is sufficient to eliminate the presence of the unwanted
divergences. In the next Section we further check this assertion
computing at two-loop order the matter four-point Green's function.

\subsection{Superpotential vertex corrections}

\begin{figure}[h!]
    \centering
    \includegraphics[width=0.5\textwidth]{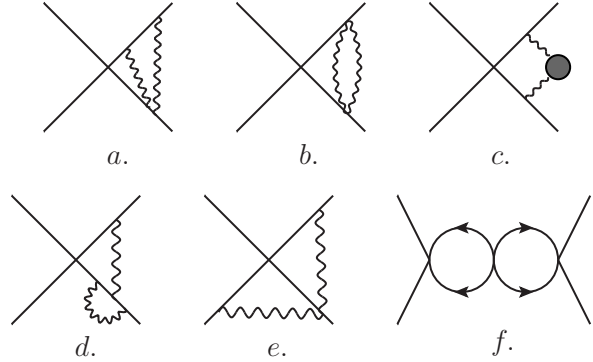}
    \caption{All two loop quantum corrections of the superpotential.}\label{1set}
\end{figure}

The set of all two-loop graphs which contribute to superpotential
corrections to leading order in $N$ in the Landau gauge are depicted
in Figure \ref{1set}; any other potentially contributing 2-loop
graph is zero due to color symmetry, supersymmetry or particular
symmetries of the Feynman integrals involved. Notice that, since in
the Landau gauge the one loop correction to the vertex is exactly
zero, we can discard many diagrams at 2-loops that contain the
1-loop diagram as a subdiagram.

To simplify notation, whenever we put a $\mathcal D(\cdots)$ in
front of the graph inside an equation, we mean the scalar graph with
all the momenta in the numerator generated after closing the
D-algebra (we put on equal foot the $V$ and $\hat V$ lines and only
in color/flavor vertex factors will we consider the sign difference
between their propagators and couplings to matter). Else, in the
absence of $\mathcal D$ in front of the graph, we just mean the
corresponding scalar Feynman integral.

To leading order in $N$, all the two loop contributions produce a
term proportional to the classical superpotential (no double traces
are generated) given by
\begin{align}\label{GeneralSuperpotentialCorrectionEquation}
&\Gamma_i[\A,\B]=\left(\frac{4\pi N}{k}\right)^2 C_i\int d^2\theta\frac{d^3p_1}{(2\pi)^3}\cdots\frac{d^3p_4}{(2\pi)^3}\ (2\pi)^3\delta\left(p_1+p_2+p_3+p_4\right)\nonumber \\
\frac{2\pi i}{k}\, \epsilon_{AC}\epsilon^{BD}
&\mathrm{tr}\left(\B^A(p_1)\A_D(p_2)\B^C(p_3)\A_B(p_4)\right)\,\mathcal{D}_i(p_1,\cdots,p_4)
\qquad i=a,\cdots,f,
\end{align}
where $C_i$ is the vertex factor of graph $i$.
$\mathcal{D}_i(p_1,\cdots,p_4)$ is the Feynman integral which
results after performing the D-algebra so as to eliminate all the
$d^4\theta$ integrals except for the last one which is used to
transform the D-operators applied on the fields into external
momenta by using that $\int d^4\theta=\int
d^2\theta\bar{D}^2(\cdots)$. The vertex factors for all graphs are
$C_a=\frac{1}{2}$, $C_b=\frac{1}{4}$, $C_c=-3$, $C_d=1$, $C_e=-1$,
$C_f=2$. We will always consider $p_1$ as the `north-western'
momentum of the graph and name the consecutive momenta
counter-clockwise.

Let us start the computation of graph $c$ of Figure \ref{1set} which
is the one we expect to give an IR divergence.  A not so
straightforward calculation of the $D$-algebra gives
\begin{align}
\mathcal{D}\left(
  \begin{picture}(35,20) (224,-35)
    \SetWidth{1.0}
    \SetColor{Black}
    \Line(225,-17)(254,-46)
    \Line(225,-47)(254,-18)
    \GOval(255,-32)(3,3)(0){0.4}
    \Photon(247,-25)(253,-30){1}{3}
    \Photon(247,-39)(252,-34){1}{3}
  \end{picture}
  \right)&=
\frac{G^{\epsilon}_{(1,1)}}{2}\int \frac{d^d k}{(2\pi)^d}
\frac{k^2(p_3+p_4)^2-{p_3}^2(k+p_4)^2-{p_4}^2(k-p_3)^2}{(k^2)^{3/2+\epsilon}(k+p_4)^2(k-p_3)^2}\nonumber \\
  &=\frac{1}{2}(p_3+p_4)^2
    \begin{picture}(33,23) (414,-30)
    \SetWidth{1.0}
    \SetColor{Black}
    \Arc[clock](419.5,-27.955)(25.521,22.957,-28.163)
    \Line(415,-13)(446,-44)
    \Line(415,-45)(446,-14)
    \Arc[clock](468.5,-28.5)(28.504,-158.385,-201.615)
  \end{picture}\
  -\frac{1}{2}\mathcal{G}_d(p_3)-\frac{1}{2}\mathcal{G}_d(p_4),
\end{align}
where we have written the Feynman integral in terms of a finite
scalar integral and infrared divergent contributions. Once again we
obtain infrared divergences when we attach a one-loop corrected
gauge vector inside a loop.  We expect that IR divergences in the
superpotential are canceled by the exact same choice for $\eta$ we
found to improve the gauge vector propagator infrared behaviour.
This is in fact true: the one-loop graph with the $\eta$-dependent
part of the gauge vector propagator gives
\begin{equation}\label{superpotential1loop}
\mathcal D\left(
\begin{picture}(34,20) (417,-147)
    \SetWidth{1.0}
    \SetColor{Black}
    \Line(416,-158)(448,-126)
    \Line(416,-126)(448,-158)
    \PhotonArc[clock](419.331,-140.449)(26.791,22.956,-27.935){0.8}{6}
    \CBox(444,-143)(448,-139){Black}{Black}
  \end{picture}
\right)=
-6\pi\lambda(\mathcal{G}_d(p_3)+\mathcal{G}_d(p_4)).
\end{equation}
With the value of the gauge parameter we made before $\eta=-6\pi
G^{\epsilon}_{(1,1)}(p^2)^{-\epsilon}\,\lambda+O\left[\lambda^3\right]$,
this insertion produces a superpotential correction with the same
structure as in (\ref{GeneralSuperpotentialCorrectionEquation}). In
this way, the sum of this graph  with graph $d$ which was also IR
divergent gives
\begin{equation}
-3(4\pi\lambda)^2\ \mathcal{D}\left(
  \begin{picture}(35,20) (224,-35)
    \SetWidth{1.0}
    \SetColor{Black}
    \Line(225,-17)(254,-46)
    \Line(225,-47)(254,-18)
    \GOval(255,-32)(3,3)(0){0.4}
    \Photon(247,-25)(253,-30){1}{3}
    \Photon(247,-39)(252,-34){1}{3}
  \end{picture}
  \right)+
4\pi\lambda\
\mathcal D\left(
\begin{picture}(34,20) (417,-147)
    \SetWidth{1.0}
    \SetColor{Black}
    \Line(416,-158)(448,-126)
    \Line(416,-126)(448,-158)
    \PhotonArc[clock](419.331,-140.449)(26.791,22.956,-27.935){0.8}{6}
    \CBox(444,-143)(448,-139){Black}{Black}
  \end{picture}
\right)=
-\frac{3}{2}(4\pi\lambda)^2\,(p_3+p_4)^2
    \begin{picture}(33,23) (414,-30)
    \SetWidth{1.0}
    \SetColor{Black}
    \Arc[clock](419.5,-27.955)(25.521,22.957,-28.163)
    \Line(415,-13)(446,-44)
    \Line(415,-45)(446,-14)
    \Arc[clock](468.5,-28.5)(28.504,-158.385,-201.615)
  \end{picture}\ ,
\end{equation}
which is finite.  These are the only dangerous IR graphs
contributing to the superpotential; the graphs which remain to be
analyzed are all finite. To show this, we list the integrals
resulting from D-algebra computations.

The simplest graph is $g$: it has three possible channels of which
two contribute to leading order in $N$. This factor of 2 is already
taken into account in the vertex factor $C_g$. The D-algebra of this
graph is simply
\begin{equation}
\mathcal{D}\left(
\begin{picture}(49,20) (415,-225)
    \SetWidth{1.0}
    \SetColor{Black}
    \Line(416,-206)(424,-222)
    \Line(416,-237)(424,-221)
    \Arc[arrow,arrowpos=0.5,arrowlength=5,arrowwidth=2,arrowinset=0.2](431,-221)(8,-0,180)
    \Arc[arrow,arrowpos=0.5,arrowlength=5,arrowwidth=2,arrowinset=0.2,clock](431,-221)(8,-0,-180)
    \Arc[arrow,arrowpos=0.5,arrowlength=5,arrowwidth=2,arrowinset=0.2,clock](447,-221)(8,-180,-360)
    \Arc[arrow,arrowpos=0.5,arrowlength=5,arrowwidth=2,arrowinset=0.2](447,-221)(8,-180,0)
    \Line(455,-221)(463,-205)
    \Line(455,-221)(463,-237)
  \end{picture}
  \right)=
\int\frac{d^3k}{(2\pi)^3}\frac{d^3l}{(2\pi)^3}\frac{-(p_1+p_2)^2}{k^2\,(k+p_1+p_2)^2\, l^2\,(l-p_3-p_4)^2}=-\frac{1}{64},
\end{equation}
In all other diagrams, a sum over different distributions of
internal lines has to be taken into account such that diagrams $b$,
$c$ and $e$ appear four times with different momentum distribution,
while diagrams $a$, $d$ appear eight times.

A straightforward calculation shows that
\begin{equation}\label{1c}
\mathcal D\left(
\begin{picture}(34,20) (417,-147)
    \SetWidth{1.0}
    \SetColor{Black}
    \Line(416,-158)(448,-126)
    \Line(416,-126)(448,-158)
    \PhotonArc(460.75,-142.25)(20.752,143.82,211.201){1}{6}
    \PhotonArc[clock](419.331,-140.449)(26.791,22.956,-27.935){0.8}{6}
  \end{picture}
\right)=
2(p_3+p_4)^2
    \begin{picture}(33,23) (414,-30)
    \SetWidth{1.0}
    \SetColor{Black}
    \Arc[clock](419.5,-27.955)(25.521,22.957,-28.163)
    \Line(415,-13)(446,-44)
    \Line(415,-45)(446,-14)
    \Arc[clock](468.5,-28.5)(28.504,-158.385,-201.615)
  \end{picture}\ .
\end{equation}
As mentioned before, this scalar integral is finite in three
dimensions. For graph $a$ we obtain the finite result
\begin{equation}
\mathcal{D}\left(
  \begin{picture}(34,20) (223,-275)
    \SetWidth{1.0}
    \SetColor{Black}
    \Line(224,-286)(256,-254)
    \Line(224,-254)(256,-286)
    \Photon(244,-265)(250,-280){1}{3}
    \Photon(252,-257)(252,-282){1}{4}
  \end{picture}
\right)=
\int \frac{d^d k}{(2\pi)^d}\frac{d^d l}{(2\pi)^d}
\frac{Tr(\gamma_{\mu}\gamma_{\nu}\gamma_{\rho}\gamma_{\sigma})\,
p_4^{\mu}\,(p_3+p_4)^{\nu}\,(k+p_4)^{\rho}\,(l-p_4)^{\sigma}}{(k+p_4)^2\,(k-p_3)^2\,(k+l)^2\,(l-p_4)^2\,l^2}.
\end{equation}
Notice that the presence of a three lined vertex is potentially
dangerous, but the momenta in the numerator of the Feynman integral
that we obtain through D-algebra guarantees finiteness. The same is
true for graph $d$
\begin{equation}
\mathcal{D}\left(
\begin{picture}(34,20) (319,-339)
    \SetWidth{1.0}
    \SetColor{Black}
    \Line(320,-317)(352,-349)
    \Line(320,-349)(352,-317)
    \Photon(346,-323)(346,-343){1.5}{5}
    \PhotonArc(343.4,-343.6)(4.808,106.928,343.072){1.5}{6.5}
  \end{picture}
  \right)=
  \int \frac{d^d k}{(2\pi)^d}\frac{d^d l}{(2\pi)^d}
\frac{-Tr(\gamma_{\mu}\gamma_{\nu}\gamma_{\rho}\gamma_{\alpha}\gamma_{\beta}\gamma_{\gamma})\,
(k+p_4)^{\mu}\,l^{\nu}\,(k+l)^{\rho}\,(k-p_3)^{\alpha}\,p_3^{\beta}\,p_4^{\gamma}}
{k^2\,(k+p_4)^2\,(k-p_3)^2\,(k+l)^2\,(l+p_3)^2\,l^2},
\end{equation}
and also for graph $e$
\begin{equation}
\mathcal{D}\left(
 \begin{picture}(34,20) (447,-146)
    \SetWidth{1.0}
    \SetColor{Black}
    \Line(448,-126)(480,-158)
    \Line(448,-158)(480,-126)
    \Photon(450,-128)(451,-154){2}{5}
    \Photon(477,-129)(451,-129){2}{5}
  \end{picture}
  \right)=
\int \frac{d^d k}{(2\pi)^d}\frac{d^d l}{(2\pi)^d}
\frac{-Tr(\gamma_{\mu}\gamma_{\nu}\gamma_{\rho}\gamma_{\sigma})\,
p_4^{\mu}\,p_2^{\nu}\,k^{\rho}\,l^{\sigma}}{k^2\,(k-p_2)^2\,(k+l+p_3)^2\,(l-p_4)^2\,l^2},
\end{equation}
which are once again finite in three dimensions.

\subsubsection{A particular exceptional momenta configuration}

By using the $\eta$ gauge fixing, we showed in the last Section that
it was possible to obtain an infrared safe function of the external
momenta for the superpotential corrections. Moreover, it is clear
that  the sum of 1PI graphs plus four-legged graphs with corrected
legs (using self-energy corrections we derived before), is a
physical gauge invariant quantity. Having found an universal finite
value for the matter propagator correction we conclude that also the
correction to the superpotential  is (at least at two loops)  gauge
independent. We would like now to compute it for a special external
momenta configuration \footnote{See \cite{Buch1} for the calculation
of the effective action on a vector superfield background.}.

The calculation we are going to present here should be interpreted
along the lines of \cite{West}. In these papers, by means of direct
computation of specific diagrams in four-dimensional supersymmetric
models, it was shown that finite contributions may survive the limit
of vanishing external momenta for the 1PI vertex function as soon as
massless particles were present.  Moreover, these
contributions could break holomorphy in the coupling constants or
supersymmetries of the action if they were to be interpreted as a
"finite renormalization" of the superpotential. It then became clear
(see \cite{Argurio} for a review and references therein) that the
correct interpretation of these contributions was to consider them
as IR singular D-terms in superspace, which are absent for instance
in the more suitable Wilsonian definition of the effective
superpotential. In what follows we would like to show that also in
the case of ABJM theories does exist a special limit of vanishing
external momenta for the vertex function which gives rise to a
finite result.

The vertex function, with the IR safe gauge choice, is guaranteed to
be finite as long as the momenta are non-exceptional. By exceptional
we mean when there exists at least one equation of the form
$\sum_{i}\rho_i\,p_i=0$ with $\rho_i$ either $0$ or $1$ and not all
$0$ nor all $1$. In our case, it is easy to see that many
exceptional configurations produce spurious IR divergences, for
example if we choose any of the four momenta, say $p_1$ to be zero.

If we were interested in finding an exceptional configuration which
is IR safe and which leads to a constant, we would need at least two
supplementary ``exceptional'' equations. In fact, we found that
modulo equivalent choices, there is only one such choice of
exceptional momenta which is IR finite. This is given by choosing
$p_1+p_2=0$ and $p_1+p_4=0$. We proceed to evaluate the graphs for
this choice.

\noindent For graph $b$ we obtain
\begin{equation}
\mathcal D\left(
\begin{picture}(34,20) (417,-147)
    \SetWidth{1.0}
    \SetColor{Black}
    \Line(416,-158)(448,-126)
    \Line(416,-126)(448,-158)
    \PhotonArc(460.75,-142.25)(20.752,143.82,211.201){1}{6}
    \PhotonArc[clock](419.331,-140.449)(26.791,22.956,-27.935){0.8}{6}
  \end{picture}
\right)=
2(p_3+p_4)^2
    \begin{picture}(33,23) (414,-30)
    \SetWidth{1.0}
    \SetColor{Black}
    \Arc[clock](419.5,-27.955)(25.521,22.957,-28.163)
    \Line(415,-13)(446,-44)
    \Line(415,-45)(446,-14)
    \Arc[clock](468.5,-28.5)(28.504,-158.385,-201.615)
  \end{picture}\longrightarrow 0.
\end{equation}
The reader might be worried that we put this graph to zero in the
exceptional configuration because of the $(p_3+p_4)^2=(p_1+p_2)^2$
numerator without taking into account that the integral multiplying
it is infrared divergent when $p_3+p_4=0$. A careful power expansion
in  $|p_3+p_4|$ gives
\begin{equation}
2(p_3+p_4)^2
    \begin{picture}(33,23) (414,-30)
    \SetWidth{1.0}
    \SetColor{Black}
    \Arc[clock](419.5,-27.955)(25.521,22.957,-28.163)
    \Line(415,-13)(446,-44)
    \Line(415,-45)(446,-14)
    \Arc[clock](468.5,-28.5)(28.504,-158.385,-201.615)
  \end{picture}=
\Bigg(\frac{1}{16\pi}\frac{\mathcal{K}\left[\sqrt{1-\frac{p_3^2}{p_4^2}}\right]}{|p_4|}\Bigg)\,|p_3+p_4|+
\Bigg(\frac{1}{16\pi^2}\frac{\log(\frac{p_3^2}{p_4^2})}{p_3^2-p_4^2}\Bigg)\,(p_3+p_4)^2+\cdots,
\end{equation}
where $\mathcal{K}(z)$ is the complete elliptic integral of the
first kind\footnote{Notice that the coefficient in front of
$|p_3+p_4|$ is implicitly symmetric under $p_3\leftrightarrow p_4$
due to the property
$\mathcal{K}\left[\sqrt{1-\frac{p_3^2}{p_4^2}}\right]=\frac{|p_4|}{|p_3|}\mathcal{K}\left[\sqrt{1-\frac{p_4^2}{p_3^2}}\right]$.
In fact, all the coefficients of the expansion have this symmetry.}
and the ellipsis are for higher orders in $|p_3+p_4|$. From this
equation we see that $p_3+p_4\rightarrow 0$ is well defined and zero
since all the coefficients in the expansion are finite in this
limit.

Graphs $a$ may be represented in terms of elementary and
Mellin-Barnes integral functions of the Lorentz invariants
$x=\tfrac{p_3^2}{(p_3+p_4)^2}$ and $y=\tfrac{p_4^2}{(p_3+p_4)^2}$
given by \footnote{Definitions, properties and relevant references
of Mellin-Barnes representation are given in the Appendix C.}
\begin{align}
&\mathcal{D}\left(
  \begin{picture}(34,20) (223,-275)
    \SetWidth{1.0}
    \SetColor{Black}
    \Line(224,-286)(256,-254)
    \Line(224,-254)(256,-286)
    \Photon(244,-265)(250,-280){1}{3}
    \Photon(252,-257)(252,-282){1}{4}
  \end{picture}
  \right)+
\mathcal{D}\left(
   \begin{picture}(34,20) (335,-258)
    \SetWidth{1.0}
    \SetColor{Black}
    \Line(336,-270)(368,-238)
    \Line(336,-238)(368,-270)
    \Photon(364,-241)(357,-259){1}{4}
    \Photon(364,-241)(364,-266){1}{4}
\end{picture}
\right)=
-(p_3+p_4)^2
    \begin{picture}(33,23) (414,-30)
    \SetWidth{1.0}
    \SetColor{Black}
    \Arc[clock](419.5,-27.955)(25.521,22.957,-28.163)
    \Line(415,-13)(446,-44)
    \Line(415,-45)(446,-14)
    \Arc[clock](468.5,-28.5)(28.504,-158.385,-201.615)
  \end{picture}\nonumber\\
&+\frac{1}{64\pi^2}\left(\tfrac{2}{3}\pi^2-Li_2(1-x)-Li_2(1-y)-\log(x)\log(y)\right)\nonumber\\
&+\frac{\sqrt{\pi}}{64\pi^3}\frac{1}{(2\pi i)^2}\int\limits_{-i\infty}^{i\infty} ds\,dt\,\Gamma^{*}(-s)\Gamma\left(\tfrac{1}{2}-s\right)\Gamma(-t)^2 \Gamma(t+s)\Gamma(1+t+s)(x^s\,y^t+x^t\,y^s).
\end{align}
This expression admits a well defined limit for $(p_3+p_4)^2\rightarrow 0$ given by
\begin{equation}
\rightarrow\frac{1}{32\pi^2}\left[\arccos^2\left(\frac{|p_3|}{|p_4|}\right)
+\arccos^2\left(\frac{|p_4|}{|p_3|}\right)+\frac{1}{4}\log^2\left(\frac{p_3^2}{p_4^2}\right)\right].
\end{equation}
If we consider more in particular that $p_3+p_4=0$, then not only
$(p_3+p_4)^2=0$ but also $p_3^2=p_4^2$, we find
\begin{equation}
\mathcal{D}\left(
  \begin{picture}(34,20) (223,-275)
    \SetWidth{1.0}
    \SetColor{Black}
    \Line(224,-286)(256,-254)
    \Line(224,-254)(256,-286)
    \Photon(244,-265)(250,-280){1}{3}
    \Photon(252,-257)(252,-282){1}{4}
  \end{picture}
  \right)+
\mathcal{D}\left(
   \begin{picture}(34,20) (335,-258)
    \SetWidth{1.0}
    \SetColor{Black}
    \Line(336,-270)(368,-238)
    \Line(336,-238)(368,-270)
    \Photon(364,-241)(357,-259){1}{4}
    \Photon(364,-241)(364,-266){1}{4}
\end{picture}
\right)
\longrightarrow 0.
\end{equation}

Now we move on to graphs $d$. By expanding the products of momenta
in the numerator of the integrals and properly completing squares
(see trace properties of $\gamma^\mu$ matrices in the appendix), one
can compare the resulting expression with the squared-completed
expression of graphs $a$ to conclude that
\begin{align}
\mathcal{D}\left(
\begin{picture}(34,20) (319,-339)
    \SetWidth{1.0}
    \SetColor{Black}
    \Line(320,-317)(352,-349)
    \Line(320,-349)(352,-317)
    \Photon(346,-323)(346,-343){1.5}{5}
    \PhotonArc(343.4,-343.6)(4.808,106.928,343.072){1.5}{6.5}
  \end{picture}
  \right)+
\mathcal{D}\left(
    \begin{picture}(34,20) (415,-336)
    \SetWidth{1.0}
    \SetColor{Black}
    \Line(416,-316)(448,-348)
    \Line(416,-348)(448,-316)
    \Photon(442,-322)(442,-342){2}{5}
    \PhotonArc[clock](440.5,-321.929)(5.369,-130.684,-326.944){2}{5.5}
    \end{picture}
    \right)&=
\mathcal{D}\left(
  \begin{picture}(34,20) (223,-275)
    \SetWidth{1.0}
    \SetColor{Black}
    \Line(224,-286)(256,-254)
    \Line(224,-254)(256,-286)
    \Photon(244,-265)(250,-280){1}{3}
    \Photon(252,-257)(252,-282){1}{4}
  \end{picture}
  \right)
+\mathcal{D}\left(
   \begin{picture}(34,20) (335,-258)
    \SetWidth{1.0}
    \SetColor{Black}
    \Line(336,-270)(368,-238)
    \Line(336,-238)(368,-270)
    \Photon(364,-241)(357,-259){1}{4}
    \Photon(364,-241)(364,-266){1}{4}
  \end{picture}
  \right)\nonumber\\
&+2(p_3+p_4)^2
    \begin{picture}(33,23) (414,-30)
    \SetWidth{1.0}
    \SetColor{Black}
    \Arc[clock](419.5,-27.955)(25.521,22.957,-28.163)
    \Line(415,-13)(446,-44)
    \Line(415,-45)(446,-14)
    \Arc[clock](468.5,-28.5)(28.504,-158.385,-201.615)
  \end{picture}
  \ -\frac{1}{32}.
\end{align}
Thus, according to the analysis we made before, in the limit $p_1+p_2=-p_3-p_4\rightarrow 0$ we obtain
\begin{equation}
\mathcal{D}\left(
\begin{picture}(34,20) (319,-339)
    \SetWidth{1.0}
    \SetColor{Black}
    \Line(320,-317)(352,-349)
    \Line(320,-349)(352,-317)
    \Photon(346,-323)(346,-343){1.5}{5}
    \PhotonArc(343.4,-343.6)(4.808,106.928,343.072){1.5}{6.5}
  \end{picture}
  \right)+
\mathcal{D}\left(
    \begin{picture}(34,20) (415,-336)
    \SetWidth{1.0}
    \SetColor{Black}
    \Line(416,-316)(448,-348)
    \Line(416,-348)(448,-316)
    \Photon(442,-322)(442,-342){2}{5}
    \PhotonArc[clock](440.5,-321.929)(5.369,-130.684,-326.944){2}{5.5}
    \end{picture}
    \right)
\longrightarrow\ -\frac{1}{32}.
\end{equation}

Finally, it is possible to calculate the Feynman integral of graph
$e$ when $p_1+p_2=0$ and $p_1+p_4=0$ by substituting $p_1=p$,
$p_2=-p$, $p_4=-p$ and from momentum conservation $p_3=p$ to obtain
\begin{equation}
\mathcal{D}\left(
 \begin{picture}(34,20) (447,-146)
    \SetWidth{1.0}
    \SetColor{Black}
    \Line(448,-126)(480,-158)
    \Line(448,-158)(480,-126)
    \Photon(450,-128)(451,-154){2}{5}
    \Photon(477,-129)(451,-129){2}{5}
  \end{picture}
  \right)\longrightarrow
\int \frac{d^d k}{(2\pi)^d}\frac{d^d l}{(2\pi)^d}
\frac{-2p^2\,k.l}{k^2\,(k+p)^2\,(k+l+p)^2\,(l+p)^2\,l^2}=\frac{1}{8\pi^2}-\frac{1}{64}.
\end{equation}
With all these elements we may make the sum to find the finite
two-loop contribution
\begin{equation}\label{2loopW}
\Gamma^{(2)}[\A,\B]=\lambda^2 (-8-\tfrac{3}{2}\pi^2)\int d^2\theta\,d^3x\
\frac{2\pi i}{k}\, \epsilon_{AC}\epsilon^{BD}
\mathrm{tr}\left(\B^A\A_D\B^C\A_B\right) .
\end{equation}
As mentioned before we expect this to be a well defined and gauge
invariant result. Nevertheless, it' s easy to show that if such
contribution had to be interpreted as a "finite renormalization" of
the superpotential it would inevitably break extended supersymmetry.

\section{General analysis} \label{section_all_loop}
We would like now to make some comments on the generality of our
results. In a three dimensional theory there are two sources of
infrared divergences in Feynman integrals. On the one hand we have
the insertion of self energy corrected lines which may produce high
powers of the propagators $\tfrac{1}{(k^2)^a}$ with
$a\ge\tfrac{3}{2}$. On the other hand the presence of a three-lined
vertex interaction with no external legs is potentially dangerous
since, if there are only scalar propagators  attached to it (no
momenta in the numerator), an IR divergence is produced after loop
integration.

\begin{figure}[h!]
    \centering
    \includegraphics[width=0.27\textwidth]{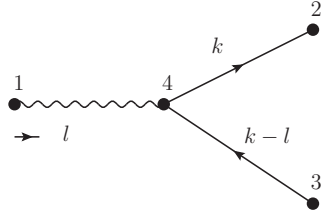}
    \caption{Matter-Gluon coupling}
    \label{matter_gluon}
\end{figure}

In general $\mathcal{N}=2$ Chern-Simons-Matter theories there are
three-lined vertexes that couple chiral fields with the gauge vector
and there is also the three gluon vertex. Consider the matter-vector
coupling as shown in figure (\ref{matter_gluon}).

If we integrate by parts on vertex $4$ at least one of the
$D$-operators of the gluon propagator, we get
\begin{equation}
\sim\frac{\bar D^{\alpha}D_{\alpha}}{l^2}\delta_{(1,4)}\,\frac{D^2\bar D^2}{k^2}\delta_{(4,2)}\,\frac{\bar D^2 D^2}{(k-l)^2}\delta_{(4,3)}=\frac{k^{\alpha\beta}}{k^2 l^2(k-l)^2}
\,D_{\alpha}\delta_{(1,4)}\,D_{\beta}\bar D^2\delta_{(4,2)}\,\bar D^2 D^2\delta_{(4,3)}.
\end{equation}
The appearance in the numerator of one of the momenta carried by the
lines eliminates the IR threat as long as no self energy corrections
are involved in the full graph (we deal with them in what follows).
A similar analysis can be done for the three gluon vertex.

It is quite obvious that the insertion of self-energy matter
corrected lines inside any given graph, does not lower the scaling
of the propagator thus not leading to IR issues. Then we conclude
that IR problems are only generated by the insertion of self-energy
corrected gluon lines: with the aid of the modified propagator we
proposed in the introduction, it seems plausible that IR divergences
can in principle be cured to all loop orders.

To leading order we have shown that the key in the elimination of IR
divergences was the completion of the 1-loop corrected gauge vector
by adding the longitudinal part with the $\eta$ piece of the
propagator. Having understood this mechanism that improves the IR
behaviour of the gauge propagator correction at 1-loop, we may
generalize this notion to all orders in $\lambda$. Due to gauge
invariance and parity we know \cite{CSComponents} that the all order
1PI vector self-energy calculated with the ordinary piece of the
propagator is given by
\begin{align}\label{all_loop_gauge_vector}
&\Delta_V=
\frac{1}{2}\int\frac{d^3k\,d^4\theta}{(2\pi)^3}\, \mathrm{Tr}\left(V(-k)
\left(\sum\limits_{l=1}A^{\epsilon}_{l}(k)\lambda^{2l}\bar{D}^{\alpha}D_{\alpha}
+\sum\limits_{l=0}B^{\epsilon}_{l}(k)\lambda^{2l+1}\,|k|\mathcal{P}_{1/2}
\right)
V(k)\right),
\end{align}
where the coefficients $A^{\epsilon}_{l}(k)$ and
$B^{\epsilon}_{l}(k)$ are functions that contain $\epsilon$-powers
of the momentum. That is, odd loop corrections contain the superspin
$1/2$ projector, and even loop corrections reproduce the original
structure of the action. Any odd-loop correction from
(\ref{all_loop_gauge_vector}), when attached inside a graph produces
a propagator given by
\begin{equation}
-\sum\limits_{l=0}B_{l}^{\epsilon}(k)\lambda^{2l+1}\,\frac{\mathcal{P}_{1/2}}{|k|}\delta_{(\theta,\theta')},
\end{equation}
which, as noted before, will produce an IR divergence. On the other
hand even-loop corrections, when attached inside a graph, produce  a
term $\sim\bar{D}^{\alpha}D_{\alpha}/k^2$  which behaves in the same
way as the basic propagator, thus not leading to IR issues. These
formulas can be readily derived using (\ref{projfo}).

Having understood the effect of corrected vector propagator
insertions in graphs, we can now proceed  to fix the $\eta$
parameter perturbatively as an odd  power series in $\lambda$. After
fixing it to order one, $\eta^{\epsilon}_{(p)}=-6\pi
G^{\epsilon}_{(1,1)}(p^2)^{-\epsilon}\lambda+O(\lambda^3)$, one
calculates every connected (not only 1PI) self energy vector
correction at order $\lambda^3$, including lower loop $O(\lambda^3)$
corrections with the $\,^1\eta^{\epsilon}_{(p)}$ piece of the
propagator. With this result we fix the next coefficient
$\,^3\eta^{\epsilon}_{(p)}$ such that we complete the transverse
projector with the longitudinal one effectively removing the source
of infrared divergence at order $\lambda^3$. This process may be
continued recursively thus improving the IR behavior of the
propagator to all loops.

In this way, if one considers a given graph which contains an
L-loop-dressed gauge vector, then if L is odd there will always be a
complementary graph in which we substitute that dressed line with
the $\eta$ piece of the propagator at the corresponding order in
$\lambda$, such that the whole line will behave as
$\sim\frac{\delta_{(\theta,\theta')}}{|k|}$; instead, when L is
even, the line behaves as the ordinary propagator
$\sim\frac{\bar{D}^{\alpha}D_{\alpha}\delta_{(\theta,\theta')}}{k^2}$
and needs no modifications. In both cases the graph will be IR safe.

Some comments are in order about the ultraviolet behaviour of the
non-locally gauged fixed theory. When we studied the effect of the
$\eta$ insertion in the case of matter self-energy graphs and
superpotential corrections, we saw that the renormalization
properties were not modified up to two loops. This makes this
alternative gauge fixing procedure consistent, since we expect any
gauge independent quantity of the theory to be independent from the
procedure. As a further non trivial check, we have also verified
that the renormalization properties of the theory in the gauge
vector sector are also not modified up to  two-loop order.

\section{Conclusions}
We studied the infrared behaviour of the off-shell amplitudes in
three-dimensional Chern-Simons-matter theories with specific
attention to the ABJM model. In $\mathcal{N}=2$ superspace IR
divergences show up in a very similar way as in four-dimensional
Super-Yang-Mills theory, being related to the corrected vector
superfield propagator insertions. At first, we showed that if the
theory is gauge fixed in a standard fashion there is no way to get
rid of the divergent integrals without losing the hermiticity of the
action. Then we  introduced a  non-local gauge fixing procedure
which leads to divergences cancelation without spoiling the
renormalizability of the theory. In order to do so, the gauge-fixing
parameter had to be perturbatively fine tuned.  Moreover, we found
in our computations  that infrared infinities seem to be always
associated to gauge dependent parts in the amplitudes,  thus not
affecting the physical quantities of the theory. As a non-trivial
output of our calculations we provided the two-loop finite
correction to the 1PI vertex function for ABJM theory in
equation (\ref{2loopW}).

It would be interesting to address the same problems in the case of
Chern-Simons-matter theories described in  $\mathcal{N}=1$
superspace, starting for instance from the formulation of BLG theory
\cite{BLG} given in \cite{MP}. In this case the analogy with the
four-dimensional case is lost and one might expect a different
infrared behaviour of superspace propagators. It would be also
interesting to perform a similar analysis in the $\mathcal{N}=3$
harmonic superspace formulation of \cite{Buch2}.

\section*{Acknowledgements}

\noindent We would like to thank A. Santambrogio, M. Siani,  C. Sieg, G. Tartaglino-Mazzucchelli and M. Wolf for useful discussions. This work has been supported in
part by INFN, PRIN prot. $2005-024045-004$.

\newpage

\appendix

\section{Superspace Notations}

We use three-dimensional superspace notations adapted from
\cite{Superspace}. We work in Euclidean space with a trivial metric
$\eta^{\mu\nu}=\eta_{\mu\nu}=\mbox{diag}(1,1,1)$ and with Dirac
Matrices $(\gamma^{\mu})_{\alpha}^{\
\beta}=i(\sigma_1,\sigma_2,\sigma_3)$. We raise and lower spinor
indexes through $\psi^{\alpha}=C^{\alpha\beta}\psi_{\beta}$ and
$\psi_{\alpha}=\psi^{\beta}C_{\beta\alpha}$, where the antisymmetric
symbols $C^{\alpha\beta}$ and $C_{\alpha\beta}$ are defined by
$C^{12}=-C_{12}=i$. Notice that with this convention contractions
are always made going from the upper left corner to the lower right
corner such that Grassmannian bilinears do not pick a sign after
hermitian conjugation:
$(\psi^2)^{\dagger}=(\psi^{\alpha}\psi_{\alpha})^{\dagger}=\bar\psi^{\alpha}\bar\psi_{\alpha}=\bar\psi^2$.
Gamma matrices satisfy
\begin{equation}
\gamma^{\mu}\gamma^{\nu}=-\eta^{\mu\nu}-\varepsilon^{\mu\nu\rho}\gamma_\rho,\quad \varepsilon^{123}=1,
\end{equation}
from which one can derive many useful trace properties such as
\begin{equation*}
Tr(\gamma^{\mu}\gamma^{\nu})=-2\eta^{\mu\nu},\quad Tr(\gamma^{\mu}\gamma^{\nu}\gamma^{\rho})=2\varepsilon^{\mu\nu\rho},
\end{equation*}
\begin{equation}
 Tr(\gamma^{\mu}\gamma^{\nu}\gamma^{\rho}\gamma^{\sigma})=2\eta^{\mu\nu}\eta^{\rho\sigma} -2\eta^{\mu\rho}\eta^{\nu\sigma} +2\eta^{\mu\sigma}\eta^{\nu\rho}.
\end{equation}
The similarity between $\mathcal N\!=\!1$ $d\!=\!4$ superspace and
$\mathcal{N}\!=\!2$ $d\!=\!3$ superspace is based on the isometry
between both isospin groups: $U(1)$ for $d\!=\!4$ and $SO(2)$ for
$d\!=\!3$. $\mathcal{N}\!=\!2$ superspace has two 2-component
anticommuting coordinates $\theta^{\alpha}_1$ and
$\theta^{\alpha}_2$ such that if one defines complex coordinates
$\theta^{\alpha}=\theta^{\alpha}_1-i\theta^{\alpha}_2$,
$\bar\theta^{\alpha}=\theta^{\alpha}_1+i\theta^{\alpha}_2$ and
complex spinor derivatives
\begin{equation}
\partial_\alpha=\frac{1}{2}(\partial_\alpha^{(1)}+i\partial_\alpha^{(2)}),\quad\quad
\bar\partial_\alpha=\frac{1}{2}(\partial_\alpha^{(1)}-i\partial_\alpha^{(2)}),
\end{equation}
they satisfy $\partial_\alpha\theta^\beta=\delta_\alpha^\beta$,
$\bar\partial_\alpha\bar\theta^\beta=\delta_\alpha^\beta$,
$\bar\partial_\alpha\theta^\beta=0$, $\partial_\alpha\bar\theta^\beta=0$.
Covariant derivatives are
\begin{equation}
D_\alpha=\frac{1}{2}(D_\alpha^{(1)}+iD_\alpha^{(2)})=
\partial_\alpha+\frac{1}{2}\bar\theta^\beta i\partial_{\alpha\beta},\quad
\bar D_\alpha=\frac{1}{2}(D_\alpha^{(1)}-iD_\alpha^{(2)})=
\bar\partial_\alpha+\frac{1}{2}\theta^\beta i\partial_{\alpha\beta},
\end{equation}
such that they carry a representation of the super-algebra
\begin{equation}\label{properties1}
\{D_\alpha,\bar D_\beta\}=i(\gamma^{\mu})_{\alpha\beta}\,\partial_{\mu}\equiv i\partial_{\alpha\beta},\quad \{D_\alpha,D_\beta\}=0,\quad\{\bar D_\alpha,\bar D_\beta\}=0.
\end{equation}
Apart from the fact that one does not make distinctions between
dotted and un-dotted spinor indexes, this is the same algebra of
covariant derivatives of $\mathcal N\!=\!1$ $d\!=\!4$ superspace
thus making Feynman supergraph rules very similar to the known
rules. On the other hand, one may construct contractions that were
not allowed in four dimensions such as $\bar D^\alpha D_\alpha$ or
$\bar\theta^\alpha\theta_\alpha$. An important property of the
vector representation is that it is symmetric:
$C^{\alpha\beta}p_{\alpha\beta}=0$ and
$C^{\alpha\beta}\partial_{\alpha\beta}=0$, which is evident after
one realizes that $\gamma^\mu$ matrices with both spinor indexes up
or down are symmetric with respect to those indexes. Defining
$\square=\partial^\mu\partial_\mu=\tfrac{1}{2}\partial^{\alpha\beta}\partial_{\alpha\beta}$,
$D^2=\tfrac{1}{2} D^{\alpha}D_\alpha$ and $\bar D^2=\tfrac{1}{2}
\bar D^{\alpha}\bar D_\alpha$ the following properties hold
\begin{equation*}
D_\alpha D^2=0,\quad \bar D_\alpha \bar D^2=0,\quad[D^{\alpha},\bar
D^2]=i\partial^{\alpha\beta}\bar D_\beta,\quad [\bar D^{\beta},
D^2]=i\partial^{\alpha\beta} D_\alpha
\end{equation*}
\begin{equation}\label{properties2}
D^2\bar D^2 D^2=\Box D^2,\quad D^{\alpha}D_\beta=\delta^\alpha_\beta D^2,\quad
\bar D^\alpha \bar D_\beta=\delta^\alpha_\beta \bar D^2.
\end{equation}
Superspin projectors are defined as
\begin{equation}
\mathcal{P}_0=\frac{1}{\square}(D^2\bar D^2+\bar D^2 D^2),\ \ \mathcal{P}_{1/2}=-\frac{1}{\square}D^{\alpha}\bar D^2 D_{\alpha},
\end{equation}
and together with $\bar D^{\alpha} D_{\alpha}$ operator, they
satisfy the useful properties
\begin{equation*}
\mathcal{P}_0^2=\mathcal{P}_0,\quad\mathcal{P}_{1/2}^2=\mathcal{P}_{1/2},\quad \mathcal{P}_0\!+\!\mathcal{P}_{1/2}=1,\quad
\mathcal{P}_0\mathcal{P}_{1/2}=0,
\end{equation*}
\begin{equation}\label{projfo}
(\bar D^{\alpha}D_{\alpha})^2=\square\mathcal{P}_{1/2},\quad \mathcal{P}_{1/2}\bar D^{\alpha}D_{\alpha}=\bar D^{\alpha}D_{\alpha},
\quad \mathcal{P}_0 \bar D^{\alpha}D_{\alpha}=0.
\end{equation}
Our conventions for integration are $\int d^2\theta=\frac{1}{2}\int d\theta^\alpha
d\theta_\alpha$, $\int d^2\bar\theta=\frac{1}{2}\int d\bar\theta^\alpha
d\bar\theta_\alpha$ and $\int d^4\theta=\int d^2\theta d^2\bar\theta$, such that
up to a total space-time derivative
\begin{equation}
\int d^2\theta\ldots =D^2\ldots|_{\theta=\bar\theta=0}\quad\mbox{and}\quad \int
d^2\bar\theta\ldots =\bar D^2\ldots|_{\theta=\bar\theta=0}.
\end{equation}
Finally, we define the $\theta$-space $\delta$-function as $\delta^4(\theta-\theta')=(\theta-\theta')^2(\bar\theta-\bar\theta')^2$.\\

\section{Feynman Rules}

We list some of the Feynman rules for ABJM theory. The vector
superfield propagators are given in the $\alpha$-gauge by:
\begin{equation}\label{trivialpropagators}
  \begin{picture}(0,0) (125,-72)
    \SetWidth{1.0}
    \SetColor{Black}
    \Photon(80,-71)(112,-71){3.5}{3}
    \Text(64,-71)[lb]{\small{\Black{$V_{\ b}^a$}}}
    \Text(112,-71)[lb]{\small{\Black{$V_{\ d}^c$}}}
  \end{picture}
=\frac{1}{p^2}\left(\bar D^{\alpha} D_{\alpha}+\alpha D^2 +\bar\alpha\bar D^2\right)\delta^4_{(\theta,\theta')}\delta^b_c\delta^d_a,
\nonumber
\end{equation}
\begin{equation}
\begin{picture}(0,0) (125,-70)
    \SetWidth{1.0}
    \SetColor{Black}
    \Photon(80,-71)(112,-71){3.5}{3}
    \Text(64,-71)[lb]{\small{\Black{$\hat V_{\ \hat b}^{\hat a}$}}}
    \Text(112,-71)[lb]{\small{\Black{$\hat V_{\ \hat d}^{\hat c}$}}}
  \end{picture}
=-\frac{1}{p^2}\left(\bar D^{\alpha} D_{\alpha}+\alpha D^2 +\bar\alpha\bar D^2\right)\delta^4_{(\theta,\theta')}\delta^{\hat b}_{\hat c}\delta^{\hat d}_{\hat a},
\end{equation}
while using the $\eta$-gauge we obtain:
\begin{equation}
  \begin{picture}(0,0) (125,-72)
    \SetWidth{1.0}
    \SetColor{Black}
    \Photon(80,-71)(112,-71){3.5}{3}
    \Text(64,-71)[lb]{\small{\Black{$V_{\ b}^a$}}}
    \Text(112,-71)[lb]{\small{\Black{$V_{\ d}^c$}}}
  \end{picture}
=\left(\frac{\bar D^{\alpha} D_{\alpha}}{p^2}+\frac{\eta_{\epsilon}(p)}{|p|}\mathcal{P}_0\right)\delta^4_{(\theta,\theta')}\delta^b_c\delta^d_a,
\nonumber
\end{equation}
\begin{equation}
\begin{picture}(0,0) (125,-70)
    \SetWidth{1.0}
    \SetColor{Black}
    \Photon(80,-71)(112,-71){3.5}{3}
    \Text(64,-71)[lb]{\small{\Black{$\hat V_{\ \hat b}^{\hat a}$}}}
    \Text(112,-71)[lb]{\small{\Black{$\hat V_{\ \hat d}^{\hat c}$}}}
  \end{picture}
=\left(-\frac{\bar D^{\alpha} D_{\alpha}}{p^2}+\frac{\eta_{\epsilon}(p)}{|p|}\mathcal{P}_0\right)\delta^4_{(\theta,\theta')}\delta^b_c\delta^d_a.
\end{equation}
From the ghost action in (\ref{fp_action}) we find the ghost propagators:
\begin{equation}
    \begin{picture}(0,0) (125,-72)
      \SetWidth{1.0}
      \SetColor{Black}
      \Line[dash,dashsize=2,arrow,arrowpos=0.5,arrowlength=3,arrowwidth=1.5,arrowinset=0.2](80,-71)(112,-71)
      \Text(64,-71)[lb]{\small{\Black{$\bar{b}_{\ b}^{\,a}$}}}
      \Text(112,-71)[lb]{\small{\Black{$c_{\ d}^{\,c}$}}}
      \end{picture}
=-\frac{1}{p^2}\delta^4(\theta-\theta')\delta^b_c\delta^d_a,
\qquad\qquad\qquad
  \begin{picture}(0,0) (125,-70) 
    \SetWidth{1.0}
    \SetColor{Black}
    \Line[dash,dashsize=2,arrow,arrowpos=0.5,arrowlength=3,arrowwidth=1.5,arrowinset=0.2](80,-71)(112,-71)
    \Text(64,-71)[lb]{\small{\Black{$\hat {\bar b}_{\ \hat b}^{\,\hat a}$}}}
    \Text(112,-71)[lb]{\small{\Black{$\hat c_{\ \hat d}^{\,\hat c}$}}}
  \end{picture}
=\frac{1}{p^2}\delta^4(\theta-\theta')\delta^{\hat b}_{\hat c}\delta^{\hat d}_{\hat a},
\end{equation}
and from $\mathcal S_{mat}$ we obtain the matter field propagators
\begin{equation}
    \begin{picture}(0,0) (125,-72)
      \SetWidth{1.0}
      \SetColor{Black}
       \Line[arrow,arrowpos=0.5,arrowlength=3,arrowwidth=1.5,arrowinset=0.2](80,-71)(112,-71)
      \Text(64,-71)[lb]{\small{\Black{$\bB_A$}}}
      \Text(115,-71)[lb]{\small{\Black{$\B^B}$}}
      \end{picture}
=\frac{1}{p^2}\delta^4(\theta-\theta')\delta^A_B\,\delta^a_b\,\delta^{\hat b}_{\hat a},
\qquad\qquad\qquad
  \begin{picture}(0,0) (125,-70) 
    \SetWidth{1.0}
    \SetColor{Black}
    \Line[arrow,arrowpos=0.5,arrowlength=3,arrowwidth=1.5,arrowinset=0.2](80,-71)(112,-71)
    \Text(64,-71)[lb]{\small{\Black{$\bA^B$}}}
    \Text(113,-71)[lb]{\small{\Black{$\A_A$}}}
  \end{picture}
=\frac{1}{p^2}\delta^4(\theta-\theta')\delta^A_B\,\delta^a_b\,\delta^{\hat b}_{\hat a},
\end{equation}
where lowercase gauge indexes are omitted if they correspond to the
same uppercase flavor index ({\it e.g.} $\bB_A\equiv(\bB_A)^{\hat
a}_{a}$, $\bA^B\equiv(\bA^B)^{b}_{\hat b}$).

Some of the non trivial color structures of the vertexes we will need are
\begin{equation}\label{colorsuperpotential1}
  \begin{picture}(135,30) (0,-10)
    \SetWidth{1.0}
    \SetColor{Black}
    \Line[arrow,arrowpos=0.5,arrowlength=4,arrowwidth=0.8,arrowinset=0.2](80,12)(96,-4)
    \Line[arrow,arrowpos=0.5,arrowlength=4,arrowwidth=0.8,arrowinset=0.2](112,12)(96,-4)
    \Line[arrow,arrowpos=0.5,arrowlength=4,arrowwidth=0.8,arrowinset=0.2](80,-20)(96,-4)
    \Line[arrow,arrowpos=0.5,arrowlength=4,arrowwidth=0.8,arrowinset=0.2](112,-20)(96,-4)
    \Text(64,5)[lb]{\normalsize{\Black{$\B^A$}}}
    \Text(114,5)[lb]{\normalsize{\Black{$\A_D$}}}
    \Text(113,-25)[lb]{\normalsize{\Black{$\B^C$}}}
    \Text(63,-25)[lb]{\normalsize{\Black{$\A_B$}}}
  \end{picture}=i\frac{4\pi}{k}{\epsilon}_{AC}{\epsilon}^{BD} \left({\delta}^{\hat{a}}_{\hat{d}}\,{\delta}^{\hat{c}}_{\hat{b}}\, {\delta}^{{d}}_{{c}}\, {\delta}^{{b}}_{{a}} - {\delta}^{\hat{c}}_{\hat{d}}\, {\delta}^{\hat{a}}_{\hat{b}}\, {\delta}^{{d}}_{{a}}\, {\delta}^{{b}}_{{c}}\right),
\end{equation}

\begin{equation}\label{colorsuperpotential2}
  \begin{picture}(135,30) (0,-10)
    \SetWidth{1.0}
    \SetColor{Black}
    \Line[arrow,arrowpos=0.5,arrowlength=4,arrowwidth=0.8,arrowinset=0.2](96,-4)(80,12)
    \Line[arrow,arrowpos=0.5,arrowlength=4,arrowwidth=0.8,arrowinset=0.2](96,-4)(112,12)
    \Line[arrow,arrowpos=0.5,arrowlength=4,arrowwidth=0.8,arrowinset=0.2](96,-4)(80,-20)
    \Line[arrow,arrowpos=0.5,arrowlength=4,arrowwidth=0.8,arrowinset=0.2](96,-4)(112,-20)
    \Text(64,5)[lb]{\normalsize{\Black{$\bB_A$}}}
    \Text(114,5)[lb]{\normalsize{\Black{$\bA^D$}}}
    \Text(113,-25)[lb]{\normalsize{\Black{$\bB_C$}}}
    \Text(63,-25)[lb]{\normalsize{\Black{$\bA^B$}}}
  \end{picture}
=i\frac{4\pi}{k}{\epsilon}^{AC} {\epsilon}_{BD}\left({\delta}_{\hat{c}}^{\hat{d}}\, {\delta}_{\hat{a}}^{\hat{b}}\, {\delta}_{{d}}^{{a}}\, {\delta}_{{b}}^{{c}}\, - {\delta}_{\hat{a}}^{\hat{d}}\, {\delta}_{\hat{c}}^{\hat{b}}\, {\delta}_{{d}}^{{c}}\, {\delta}_{{b}}^{{a}}\right),
\end{equation}

\begin{equation}\label{2vectormatter}
   \begin{picture}(100,30) (185,-150) 
    \SetWidth{1.0}
    \SetColor{Black}
    \Line(192,-153)(256,-153)
    \Photon(192,-121)(224,-153){3}{5}
    \Photon(224,-153)(256,-121){3}{5}
    \Text(177,-130)[lb]{\normalsize{\Black{$V^b_c$}}}
    \Text(260,-133)[lb]{\normalsize{\Black{$V^d_f$}}}
    \Text(260,-153)[lb]{\normalsize{\Black{$\B^E$}}}
    \Text(176,-153)[lb]{\normalsize{\Black{$\bB_A$}}}
  \end{picture}
=\frac{1}{2}\frac{4\pi}{k}\delta^A_E\,\delta^{\hat e}_{\hat a}\,
\left(\delta^a_b\,\delta^c_d\,\delta^f_e+\delta^a_d\,\delta^f_b\,\delta^c_e\right),
\end{equation}
\begin{equation}
   \begin{picture}(70,50) (360,-30)
    \SetWidth{1.0}
    \SetColor{Black}
    \Photon(352,22)(384,-10){3}{5}
    \Photon(384,-10)(416,22){3}{5}
    \Photon(384,-10)(384,-42){3}{4}
    \Text(332,11)[lb]{\normalsize{\Black{$V^c_{\ d}$}}}
    \Text(422,8)[lb]{\normalsize{\Black{$V^e_{\ f}$}}}
    \Text(392,-42)[lb]{\normalsize{\Black{$V^a_{\ b}$}}}
    \Line[arrow,arrowpos=0.5,arrowlength=5,arrowwidth=2,arrowinset=0.2](368,-42)(368,-26)
    \Line[arrow,arrowpos=0.5,arrowlength=5,arrowwidth=2,arrowinset=0.2](368,-10)(352,6)
    \Line[arrow,arrowpos=0.5,arrowlength=5,arrowwidth=2,arrowinset=0.2](400,-10)(416,6)
    \Text(364,-46)[lb]{\small{\Black{$\rput[lb]{90}{p+q}$}}}
    \Text(416,-10)[lb]{\small{\Black{$\rput[lb]{45}{p}$}}}
    \Text(352,-10)[lb]{\small{\Black{$\rput[lb]{-45}{q}$}}}
  \end{picture}=
\frac{1}{2}\sqrt{\frac{4\pi}{k}}\left(\delta^{b}_{c}\,\delta^d_e\,\delta^f_a-\delta^b_e\,\delta^f_c\, \delta^d_a\right)\left[D^{\alpha}(q)\bar D_{\alpha}(p)-\bar D^{\alpha}(q)D_{\alpha}(p)\right],
\end{equation}
\\
\noindent and similarly for other vertexes involving $\A$ and $\hat
V$ fields. We apply the usual D-algebra rules and regularize
integrals when needed using dimensional reduction prescriptions.

\section{Relevant Integrals}
In the computation of two-point functions we introduced the function
$G^{\epsilon}_{(a,b)}$ defined as
\begin{equation}
\int\frac{d^d k}{(2\pi)^d}\frac{1}{k^{2a}\ (k+p)^{2b}}=\frac{G^{\epsilon}_{(a,b)}}{\left(p^2\right)^{a+b-d/2}}=\frac{\Gamma(a+b-d/2)\Gamma(d/2-a)\Gamma(d/2-b)} {(4\pi)^{d/2}\Gamma(a)\Gamma(b)\Gamma(d-a-b)\left(p^2\right)^{a+b-d/2}},
\end{equation}
and a particular two loop infrared divergent integral
\begin{align}\label{infrareddivergentfunction}
\mathcal{G}_d(p)&=\int\frac{d^d k\, d^d l}{(2\pi)^{2d}}\frac{p^2}{l^2\,(l+k)^2\,(k+p)^2\,k^2}=
G^{\epsilon}_{(1,1)}G^{\epsilon}_{(1,3/2+\epsilon)}(p^2)^{-2\epsilon} \\
&=(p^2)^{-2\epsilon}\left(-\frac{1}{64 \pi ^2\,\epsilon}+\frac{1+\gamma-\log(4\pi )}{32 \pi ^2}+\mathcal{O}(\epsilon)\right).
\end{align}
All the relevant integrals in the calculation of the propagators can
be reduced to the $G$-function form.

In the computation of four-point integrals in the exceptional
configuration, we found it necessary to expand Feynman integrals in
powers of the kinematic invariants in order to carefully take the
appropriate limits. In the end, the correct limit through this
analysis became a confirmation of the ``naive'' result since it
coincides with the limit taken directly on the integrand. To deal
with the expansions we used multiple Mellin-Barnes contour
representation of vertex integrals \cite{Davydychev,Smirnov}. These
representations are based on the identity
\begin{equation}\label{0F1}
\frac{1}{(k^2+M^2)^a}=\frac{1}{(M^2)^a\Gamma(a)}\,\frac{1}{2\pi i}\int\limits_{-i\infty}^{i\infty} ds\,\Gamma(-s)\Gamma(s+a)\left(\frac{k^2}{M^2}\right)^s,
\end{equation}
where the contour is given by a straight line along the imaginary
axis such that indentations are used if necessary in order to leave
the series of poles $s=0,1,\cdots,n$ to the right of the contour and
the series $s=-a,-a-1,\cdots,-a-n$ to the left of the contour. After
Feynman-parametrizing a triangle integral and using ($\ref{0F1}$),
the following formula holds
\begin{equation*}
\int\frac{d^d k}{(2\pi)^d}\frac{1}{k^{2\mu_1}(k-p)^{2\mu_2}(k+q)^{2\mu_3}}= \frac{(4\pi)^{-d/2}(p+q)^{2(d/2-\sum_i\mu_i)}}{\prod_i\Gamma(\mu_i)\Gamma(d-\sum_i\mu_i)}\times
\end{equation*}
\begin{equation*}
\times\left(\frac{1}{2\pi i}\right)^2\int\limits_{-i\infty}^{i\infty}ds\,dt\,\Gamma(-s)\Gamma(-t)\Gamma(\tfrac{d}{2}-\mu_1-\mu_2-s)\Gamma(\tfrac{d}{2}-\mu_1-\mu_3-t)\times
\end{equation*}
\begin{equation}
\Gamma(\mu_1+s+t)\Gamma({\sum}_i\mu_i-\tfrac{d}{2}+s+t)\left(\frac{p^2}{(p+q)^2}\right)^s\left(\frac{q^2}{(p+q)^2}\right)^t;
\end{equation}
and for a vector-like triangle we have
\begin{equation*}
\int\frac{d^d k}{(2\pi)^d}\frac{k^{\nu}}{k^{2\mu_1}(k-p)^{2\mu_2}(k+q)^{2\mu_3}}= \frac{(4\pi)^{-d/2}(p+q)^{2(d/2-\sum_i \mu_i)}}{\prod_i\Gamma(\mu_i)\Gamma(d-\sum_i\mu_i+1)}\left(\frac{1}{2\pi i}\right)^2\times
\end{equation*}
\begin{equation*}
\times\int\limits_{-i\infty}^{i\infty}ds\,dt\, \Gamma(-s)\Gamma(-t)\Gamma(\mu_1+s+t)\Gamma({\sum}_i\mu_i-\tfrac{d}{2}+s+t) \left(\frac{p^2}{(p+q)^2}\right)^s\left(\frac{q^2}{(p+q)^2}\right)^t\times
\end{equation*}
\begin{equation}
\left[\Gamma(\tfrac{d}{2}\!-\!\mu_1\!-\!\mu_2\!-\!s)\Gamma(\tfrac{d}{2}\!-\!\mu_1\!-\!\mu_3\!-\!t\!+\!1)p^{\,\nu}\! -\!\Gamma(\tfrac{d}{2}\!-\!\mu_1\!-\!\mu_2\!-\!s\!+\!1)\Gamma(\!\tfrac{d}{2}\!-\!\mu_1\!-\!\mu_3\!-\!t)q^{\,\nu}\right],
\end{equation}
where the multiple contours are taken using the same convention as
the first definition unless otherwise indicated. It is customary to
indicate with a $^*$ over the $\Gamma(z)$ function the case where
one leaves a pole to the other of the conventional side of the
contour. With these representations, among with Barnes 1$^{st}$ and
2$^{nd}$ lemmas
\begin{equation}
\frac{1}{2\pi i}\int ds\, \Gamma(a+s)\Gamma(b+s)\Gamma(c-s)\Gamma(d-s)=\frac{\Gamma(a+c)\Gamma(a+d)\Gamma(b+c)\Gamma(b+d)}{\Gamma(a+b+c+d)},
\end{equation}
\begin{align}
&\frac{1}{2\pi i}\int ds\, \frac{\Gamma(a+s)\Gamma(b+s)\Gamma(c+s)\Gamma(d-s)\Gamma(-s)}{\Gamma(e+s)}\nonumber\\
&=\frac{\Gamma(a)\Gamma(b)\Gamma(c)\Gamma(a+d)\Gamma(b+d)\Gamma(c+d)}{\Gamma(e-a)\Gamma(e-b)\Gamma(e-c)} ,\quad\mbox{with}\quad e=a+b+c+d,
\end{align}
and their multiple corollaries, we were able to carefully expand the
2-loop four-point integrals in the relevant kinematic invariants in
order to take the limit of exceptional momenta.

\end{document}